\documentclass{epl}

\usepackage{psfig,epsfig,array}

\title{Possible Gigantic Variations on the Width of Viscoelastic Fingers}
\author{Eugenia Corvera Poir\'e\inst{1} \and J. Antonio del R\'{\i}o\inst{2}}

\institute{
  \inst{1} Departamento de F\'{\i}sica y Qu\'{\i}mica Te\'orica,
Facultad de Qu\'{\i}mica, UNAM. Ciudad Universitaria,
M\'exico, D.F. 04510, M\'exico.\\
  \inst{2} Centro de Investigaci\'on en Energ\'{i}a, UNAM.  AP 34,
62580 Temixco, Morelos}

\begin{document}

\maketitle

\begin{abstract}
We analyze the effect of frequency on the width of a single
finger displacing a viscoelastic fluid. We derive a generalized
Darcy's law in the frequency domain for a linear viscoelastic
fluid flowing in a Hele Shaw cell. This leads to an analytic expression
for the dynamic permeability that has maxima which are several
orders of magnitude larger than the static permeability.
We then follow an argument of de Gennes~\cite{degennes} to
obtain the smallest possible finger width when viscoelasticity is important.
Using this, and a conservation law, we obtain a lowest bound for the width
of a single finger displacing a viscoelastic fluid. 
Our results indicate that when a small amplitude signal of the frequency 
that maximizes the permeability is overimposed to a constant pressure drop, 
gigantic variations are obtained for the finger width.
\end{abstract}

Many problems associated with oil, plastic, and chemical industries involve
the response of a fluid in a porous medium to a frequency dependent pressure
drop. Laboratory experiments have shown for example that ultrasonic
vibrations can considerably increase the rate of flow of a liquid through a
porous medium \cite{expe}. A useful way to describe such frequency-dependent
processes is through a dynamic permeability function \cite{todos}.
Expressions for the dynamic permeability of Newtonian fluids were obtained
long ago for deformable and isotropic porous media \cite{ping}. More
recently, an enhancement of several orders of magnitude in the dynamic
permeability was found for the case of Maxwellian fluids \cite{delrio}.  
Many other problems in the same industries, involve the displacement of a
viscoelastic fluid by a second fluid of lower viscosity. The dramatic
enhancement found for the dynamic permeability of a Maxwell fluid \cite
{delrio} flowing in a tube, immediately suggests that similar dramatic
effects might happen in problems of flow involving more than one fluid.

The viscous fingering problem has been historically very important in the
area of morphology of interfaces out of equilibrium~\cite{patternform} and
has been a model system to describe displacement of viscous fluids in porous
media. The viscous fingering problem is the prediction of the shape of the
fluid interface in a two-phase flow confined in a Hele-Shaw cell~\cite{hele}
which consists of a pair of glass plates parallel to each other separated by
a small gap. A viscous fluid occupies the space between the plates and it is
pushed by a second fluid whose viscosity is relatively low. When the fluid
is pushed laterally, the experiment is said to take place in a linear cell.
When the fluid is injected through a hole made in one of the plates forming
the cell, the experiment is said to take place in a radial cell. Both the
linear and the radial geometry have been studied extensively. The interface
between the fluids is unstable and the structures that are formed when the
interface destabilizes are called fingers. This is the so called
Saffman-Taylor instability~\cite{saffman}. Recently, viscous fingering
experiments have been made with non-Newtonian fluids. Some of these
experiments have used clays~\cite{vandamme}, polymer solutions~\cite
{maher,bonn} and lyotropic lamellar phases~\cite{kellay}. Non-Newtonian
fluids differ widely in their physical properties, with different fluids
exhibiting a range of different properties, from plasticity and elasticity
to shear thickening and shear thinning. Several studies have been made, both
theoretically and experimentally in order to differentiate between the
effects caused by different properties ~\cite{degennes,bonn,shelley,corvera}.

In this letter we analyze the effect of frequency on the width of a single
finger displacing a viscoelastic fluid. To start, we derive a generalized
Darcy's law in the frequency domain for a linear viscoelastic fluid flowing
in a Hele Shaw cell. This leads to an analytic expression for the dynamic
permeability that in agreement with results in other geometries~\cite{delrio}
has maxima which are several orders of magnitude larger than the static
permeability. We then follow an argument of de Gennes~\cite{degennes} to
obtain the smallest possible finger width when viscoelasticity is important.
Using this, and a conservation law, we obtain a lowest bound for the width
of a single finger displacing a viscoelastic fluid. What we find is that,
when there is a small amplitude signal of the frequency that maximizes
the permeability, overimposed to a constant pressure drop,
gigantic variations are obtained for the finger width,
that is, tiny perturbations in the pressure drop
produce dramatic effects on the amplitude of the finger width.

We start our study taking a Maxwell fluid, which is the simplest model of a
linear viscoelastic fluid, and linearize\ the equation governing the flow.
In the frequency domain this equation is\cite{delrio} 
\begin{equation}
-\rho \left( t_{r}\omega ^{2}+i\omega \right) \vec{v}-\eta \nabla ^{2}\vec{v}%
=-\left( 1-i\omega t_{r}\right) \nabla p\,\,\,\,\,\,\,\,\,\,.  \label{buena}
\end{equation}
Here both the velocity $\vec{v}$ and the pressure $p$ are in the frequency
domain, that is, they are functions of space and frequency. $t_{r}$, $\eta$
and $\rho$ are respectively the relaxation time of the Maxwell model and the
viscosity and the density of the fluid.
We solve (\ref{buena}) for a homogeneous fluid
flow in the $x$ direction, confined between parallel plates at $z=\pm l$
subject to the boundary conditions $v_{x}(\pm l)=0$. We obtain the velocity
profile between the plates. In order to obtain a generalized Darcy's law, we
average over the $z-$direction and obtain for the average flow 
\begin{equation}
\left\langle v\right\rangle =\left( 1-\frac{\tan \sqrt{\beta }l}{\sqrt{\beta 
}l}\right) \frac{\left( 1-i\omega t_{r}\right) }{\beta \eta }\frac{dp}{dx}%
\,\,\,\,\,\,\,\,\,\,.
\end{equation}
Here $\beta (\omega )=\frac{\rho \left( t_{r}\omega ^{2}+i\omega \right) }{%
\eta }$. In the limit $\omega \rightarrow 0$ we recover the steady state
Darcy's law and for $t_{r}\rightarrow 0$ we obtain the newtonian fluid case.
The dynamic permeability of a single Hele-Shaw cell is then given by 
\begin{equation}
K\left( \omega \right) =-\left( 1-\frac{\tan \sqrt{\beta }l}{\sqrt{\beta }l}%
\right) \frac{\left( 1-i\omega t_{r}\right) }{\beta }\,\,\,\,\,\,\,\,\,\,.
\label{dynperm}
\end{equation}

\begin{figure}[]
\begin{center}
\resizebox{!}{7cm}{\includegraphics{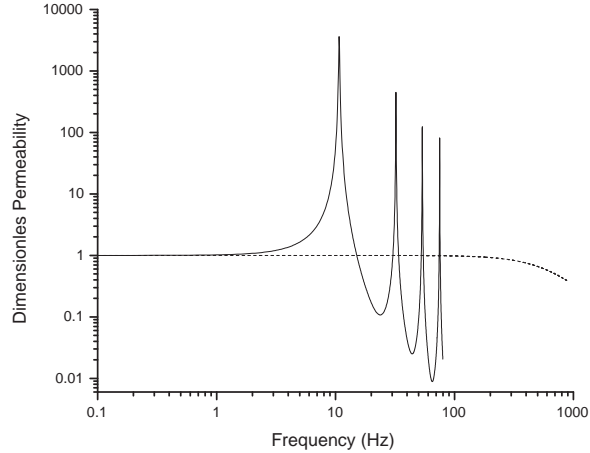}}
\end{center}
\caption{Dimensionless Dynamic Permeability vs Frequency. The continuous
line is for a Viscoelastic Fluid with a relaxation time $t_r=6s$, the dotted
line for a Newtonian fluid. For both lines the viscosity $\protect\eta=0.7 p$%
, the density $\protect\rho=1g/cm^3$ and the spacing between the plates $%
L=2mm$}
\label{maxvsnewt}
\end{figure}

\begin{figure}[]
\begin{center}
\resizebox{!}{5cm}{\includegraphics{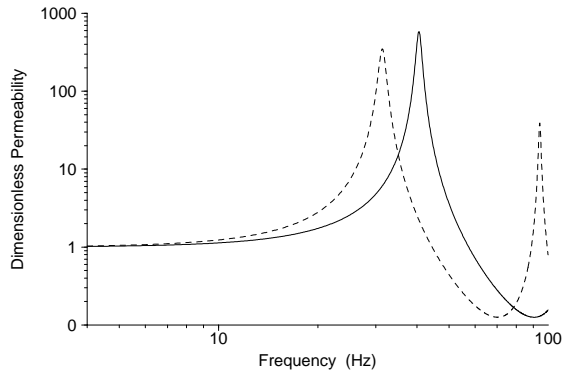}}
\end{center}
\caption{Effect of viscosity on the dimensionless dynamic permeability for a
viscoelastic fluid. For the dotted line the viscosity $\protect\eta=0.6 p$,
for the continuous line the viscosity $\protect\eta=1 p$. For both lines the
density $\protect\rho=1g/cm^3$, the spacing between the plates $L=1mm$ and
the relaxation time $t_r=0.6s$}
\label{maxvisc}
\end{figure}

Figure \ref{maxvsnewt} shows the real part of the normalized permeability $%
K(\omega )/K(0)$ versus the frequency for both a Maxwell fluid and a
Newtonian fluid. The figure shows that for a Newtonian fluid the
permeability is a monotonically decreasing function of frequency. For a
Maxwell fluid on the other hand, there are frequencies in which there are
resonances which give peaks for the permeability. For typical values of
density, viscosity, plate spacing and relaxation time in the common
literature of Hele-Shaw problems, we find that this permeability can be two
or three orders of magnitude larger than the Newtonian permeability, this
is because the behavior of the flow at such frequencies is dominated by the
elastic properties of the fluid. What this result means is that if we
displace the viscous fluid at the frequency that maximizes the permeability,
the fluid will flow with the least possible resistance. We find that the
frequency which gives the largest permeability is of the order of ten Hertz.

Figure \ref{maxvisc} shows how the maximum of the real part of the
permeability shifts towards higher frequencies when the viscosity is
increased. So, the more viscous the fluid, the higher the frequency of the
necessary pumping to minimize the resistance to flow.

We now analyze the problem of a low viscosity fluid displacing a high
viscosity viscoelastic fluid. In particular, we analyze the case of a single
finger with a time dependent velocity $U(t)$ propagating into the viscous
fluid. We call $\lambda (t)$ the finger width. It is worth emphasizing that
we are not considering a steady state. Both $U$ and $\lambda $ depend on
time \footnote{%
We have recently shown that the single finger solution exists for this case.
Whether or not the solution is stable has not been shown.}.

The quantity $U/\lambda $ gives a characteristic frequency. The viscoelastic
fluid has a characteristic time $t_{r}$ which in the case of a Maxwell fluid
is given by $t_{r}=\eta /G$, $G$ being the rigidity modulus of the viscous
fluid. When $U/\lambda >1/t_{r}$ the viscoelastic fluid behaves like a solid
and there is no Saffman-Taylor instability. The allowed wavelengths should
all correspond to $U/\lambda <1/t_{r}$~\cite{degennes}. Therefore, the
smallest possible finger width should be such that 
\begin{equation}
\frac{U(t)}{\lambda (t)}=\frac{1}{t_{r}}\,\,\,\,\,\,\,\,\,\,.
\label{argdegennes}
\end{equation}
Conservation of matter implies that 
\begin{equation}
U(t)\lambda (t)=V_{\infty }(t)\,\,\,\,\,\,\,\,\,\,.  \label{conserv}
\end{equation}
From equations (\ref{argdegennes}) and (\ref{conserv}) we can relate the
finger width $\lambda (t)$ and the tip velocity $U(t)$ with the velocity at
the extreme of the cell $V_{\infty }(t)$ as 
\begin{equation}
\lambda ^{2}(t)=\frac{t_{r}}{W}V_{\infty }(t)\,\,\,\,\,\,\,\,\,\,.
\label{lambda}
\end{equation}
and 
\begin{equation}
U^{2}(t)=\frac{W}{t_{r}}V_{\infty }(t)\,\,\,\,\,\,\,\,\,\,.  \label{tipvel}
\end{equation}

Experimentally, the parameter that can be controlled is the pressure
difference at the extremes of the cell. So, given $\Delta p(t)$, we can make
the following calculations: 
\begin{eqnarray}
\Delta p(t) & \rightarrow & \Delta \hat{p}(\omega),  \nonumber \\
& Fourier \,\,\,\, Transform &  \nonumber
\end{eqnarray}
\begin{eqnarray}
\Delta \hat{p}(\omega) & \rightarrow & \hat{V}_\infty(\omega),  \nonumber \\
& Darcy^{\prime}s \,\,\,\, law \,\,\,\, in \,\,\,\, frequency \,\,\,\,
domain &  \nonumber
\end{eqnarray}
\begin{eqnarray}
\hat{V}_\infty(\omega) & \rightarrow & {V}_\infty(t),  \nonumber \\
& Inverse \,\,\,\, F.T. &  \nonumber
\end{eqnarray}
\begin{eqnarray}
{V}_\infty(t) & \rightarrow & \lambda(t), U(t)  \nonumber
\end{eqnarray}

In order for fingers to exist, the pressure difference should be at any
moment positive. Simple oscillatory signals are not possible since the
instability will exist for only half of the period. So we are thinking, for
example, in signals that are overimposed to pressure drops that are large
enough to destabilize the interface.

We consider the simple case of an oscillatory finger. Suppose we impose a
pressure difference of the form 
\begin{equation}
\Delta p(t)=p_{o}+p_{a}e^{-i\omega _{0}t},
\end{equation}
that is, a constant pressure drop plus an oscillatory signal of frequency $%
\omega _{0}$. We obtain the following expressions for the different steps
presented above: 
\begin{equation}
\Delta \hat{p}(\omega )=\sqrt{2\pi }p_{o}\delta (\omega )+\sqrt{2\pi }%
p_{a}\delta (\omega -\omega _{0})
\end{equation}
\begin{equation}
\hat{V}_{\infty }(\omega )=\frac{\sqrt{2\pi }}{\eta L}K(\omega )\left[
p_{o}\delta (\omega )+p_{a}\delta (\omega -\omega _{0})\right] 
\end{equation}
\begin{equation}
{V}_{\infty }(t)=\frac{p_{o}}{\eta L}K(0)+\frac{p_{a}}{\eta L}K(\omega
_{0})e^{-i\omega _{0}t}
\end{equation}
Now, for the steady state we know that 
\begin{equation}
{V}_{\infty }^{ss}=\frac{p_{o}}{\eta L}K(0)
\end{equation}
and then we can express all of our results as 
\begin{equation}
{V}_{\infty }(t)={V}_{\infty }^{ss}\left[ 1+\frac{p_{a}}{p_{0}}\frac{%
K(\omega _{0})}{K(0)}e^{-i\omega _{0}t}\right] ,
\end{equation}
\begin{equation}
\lambda ^{2}(t)=\lambda _{ss}^{2}\left[ 1+\frac{p_{a}}{p_{0}}\frac{K(\omega
_{0})}{K(0)}e^{-i\omega _{0}t}\right] ,
\end{equation}
\begin{equation}
U^{2}(t)=U_{ss}^{2}\left[ 1+\frac{p_{a}}{p_{0}}\frac{K(\omega _{0})}{K(0)}%
e^{-i\omega _{0}t}\right] .
\end{equation}
Here $\lambda _{ss}$ and $U_{ss}$ are defined in terms of ${V}_{\infty }^{ss}
$ trough (\ref{lambda}) and (\ref{tipvel}). What it is interesting to note, is
that the behavior is totally different if the displaced viscous fluid is
Newtonian or viscoelastic. For Newtonian fluids, the ratio $\frac{K(\omega
_{0})}{K(0)}$ is always small or equal to one and the variations in the
width of the finger are small. In the other hand, for viscoelastic fluids,
the ratio $\frac{K(\omega _{0})}{K(0)}$ can be several orders of magnitude
larger than in the Newtonian case. Hence, if the imposed signal has a
frequency $\omega _{0}$ such that it maximizes the dynamic permeability, and
therefore the ratio $\frac{K(\omega _{0})}{K(0)}$, a small amplitude in the
oscillatory part of the
pressure signal, will result in gigantic variations in the amplitude of
both the finger velocity and the finger width.

A word of caution is needed, since when one does not consider surface
tension the finger width and the velocity at the finger tip are not
independent. It is worth noticing that since the surface tension has a
stabilizing effect for small wavelengths any consideration of surface
tension would give fingers wider or equal to the lowest bound that we are
dealing with in the present work.

We are not solving the selection problem, in order to do it and to compute
finger shapes, the problem should be solved as in references~\cite
{saffman,mclean}. Nevertheless the main result of the present paper should
remain. That is, small variations in the pressure drop at a frequency that
maximizes the permeability, will result in gigantic variations of the finger
width.

\acknowledgments This project was partially supported by grants 
CONACYT 33920-E and
38538-E. We thank Mariano L\'{o}pez de Haro for a critical reading of the
manuscript and enlightening discussions.

\end{document}